\documentclass[preprint,amsmath,amssymb,aps,showkeys,showpacs,pra,]{revtex4-1}

\usepackage[english]{babel}
\usepackage{graphicx}  
\usepackage{amsmath}
\usepackage{dcolumn}
\usepackage{amssymb}
\usepackage{bm}
\usepackage[latin1]{inputenc}
\usepackage{graphicx,color}
\usepackage[pdftex]{hyperref}

\begin{document}
\title{Quantum Ring in Gapped Graphene Layer with Wedge Disclination in the Presence of an Uniform Magnetic Field }
\author{José Amaro Neto, J. R. de S. Oliveira, Claudio Furtado 
\email{furtado@fisica.ufpb.br}
and Sergei Sergeenkov}
\affiliation{Departamento de F\'isica, Universidade Federal da Para\'iba, Caixa Postal 5008, 58051-970, Jo\~ao Pessoa, PB, Brazil.} 

\begin{abstract}
In this paper we investigate the relativistic quantum dynamics of  a massive excitation in a  graphene  layer with a wedge disclination in the presence of an uniform magnetic field.  We use a Dirac oscillator type coupling to introduce the confining potential  for massive fermions  in this system.  We obtain the energy spectrum  and  eigenfunctions for the quantum ring pierced by Aharonov-Bohm flux resulting in appearance of persistent current and spontaneous magnetization. 

\end{abstract}


\maketitle
\section{Introduction}\label{sec1}
In recent  years, some important and interesting physical properties of quantum dots in graphene under the influence of applied magnetic field have attracted much attention \cite{pono, schn1}. Among numerous theoretical studies, it is worthwhile to mention recent results on different properties of quasiparticles confined in nanostructures, including  quantum dots \cite{silv, schn2}   and quantum rings \cite{peters1,peters2,peters3}. In paricular, it was demonstrated that electronic structure, magnetic and transport properties can be significantly altered for graphene in the presence of disclinations. Of special interest is the low-energy behavior of quasiparticles in graphene. At low energies, the quantum behavior of excitations in graphene is  described by an equation analogue to a  massless Dirac equation \cite{vozmediano,castro}.  In the absence of  a sublattice symmetry, a gap appears and the corresponding dynamics can be incorporated into  a massive term in the Dirac equation.   It is a known fact that a band gap is induced in these samples, and massive Dirac fermions play a central role in the low-energy limit \cite{Zhou}.  It was observed \cite{Zhou} that a substrate induced potential can break the chiral symmetry  existing in graphene and, in this way, introduce a mass term in the spectrum   of  quasiparticles in  graphene. In general, this case is called gapped graphene.   Recently \cite{6,7,8,9},  it was demonstrated that the presence of impurities  breaks partially the  symmetries of the honeycomb lattice in   graphene and  generates a Dirac mass for the excitations in this material. In this way, the  excitations in the gapped graphene are described by a massive Dirac equation.
Several studies  on the influence of  various impurities, electron-electron interactions and  substrate structures  responsible for appearance of a mass gap in graphene have been carried out \cite{33,35,36,38,39,40,41,44,46,47,50}. These effects   imply breaking of sublattice symmetry leading to a mass gap in a graphene layer.

The Hamiltonian  describing the $\pi$  orbitals in a gapped graphene is given by
\begin{eqnarray}
\mathcal{H}=  t \sum_{i=A} \sum_{j=1}^{3}\left[a^{\dagger}(\vec{r_{i}})b(\vec{r_{i}}+\vec{u_{j}})+b^{\dagger}(\vec{r_{i}}+\vec{u_{j}})a(\vec{r_{i}})\right] +\\ \nonumber \beta\sum_{i=A}\left[a^{\dagger}(\vec{r_{i}})a(\vec{r_{i}})-b^{\dagger}(\vec{r_{i}}+\vec{u_{j}})b(\vec{r_{i}}+\vec{u_{j}})\right],
\end{eqnarray}
where $t$ is the probability of transition by tunneling, and  $2\beta$ is the energy difference of electrons on sites $A$ and $B$. For massless graphene $\beta=0$. The creation and annihilation operators create and annihilate electrons in their respective sublattices, i.e., $a$ and $a^{\dagger}$ act  in the sublattice $A$, while $b$ and $b^{\dagger}$  in the sublattice $B$. The vectors $\vec{u_1}$, $\vec{u_2}$ and $\vec{u_3}$ connect the sites $A$ to  their first neighbors. 

We can solve the eigenvalue problem by diagonalizing the matrix inside the sum \cite{Chakraborty,jose}. The corresponding eigenvalues are:
\begin{eqnarray}
E=\pm \left( \beta^{2} + t^{2}\left[3+2\cos\left(\sqrt{3}k_{y}d\right)+4\cos\Biggl(\frac{3}{2}k_{x}d\Biggl)\cos\Biggl(\frac{\sqrt{3}}{2}k_{y}d\Biggl) \right]\right)^{1/2},
\end{eqnarray}  
where $d$ is the distance between carbon atoms in the lattice. The  energy  eigenvalues display minima in the corners of the Brillouin zone.  In this case the separation between the two bands assumes a minimal  value at the corners. Notice that only two of six points of the corners of  hexagonal Brillouin zone are inequivalent. Therefore, these two points are threefold degenerate.  This degeneracy occurs at six points located at the vertices of the first Brillouin zone in the form of a hexagon. Each of them forms, separately, the Fermi surface of graphene, and therefore they are called the Fermi points. In the present case these  six points are reduced to only two, which we call $\vec{K_{+}}$ and $\vec{K_{-}}$. 

Now let us consider a continuum approximation for small momenta near $K_{\pm}$,  where electrons are found to show a Dirac-like behavior. In this limit, the Hamiltonian assumes the following form
\begin{equation}
\hat{H}=\int \Psi^{\dagger} H\Psi d^{2}r
\end{equation} 
Here, $H$ is the corresponding one-particle Weyl-Dirac Hamiltonian given by
\begin{eqnarray}\label{hamilton}
H=\left(-i\gamma^{i}\partial_{i} +\gamma^{0}m\right),
\end{eqnarray}
where $\gamma^{\mu}$ are the gamma matrices and the spinor is given by 	$\Psi^{T} = (\psi_a, \psi_b, \chi _a, \chi _b)$ 
with $a$ and $b$ being the indices related to  $A$ and $B$ sublattice; the spinors $\psi$ and  $\chi$  correspond  to the $K_{\pm}$ Fermi points.

The role  of the mass term in graphene is very important for some electronic  applications \cite{vozmedianoreview}. From the viewpoint of the quantum field theory,  generation of an electron mass or opening of a gap in graphene is protected by three-dimensional version of chiral symmetry. It means that  we can not open a gap in this material by simply accounting for radiative corrections. A possibility to open a gap in graphene spectrum based on breaking of the product of time reversal  and inversion symmetry under an external magnetic field (or when the two Fermi points are involved in case  of the Kekule distortion) have been discussed in Refs. \cite{33,35,36,38,39,40,41,44,46,47,50,vozmedianomanes}. Another possibility based on breaking  of  the chiral symmetry  was considered for the case of Haldane model \cite{haldane} and Kane-Mele model \cite{35} for topological insulators.

 It is well known from theoretical and experimental studies  of nanostructures of graphene that for large enough applied magnetic fields $B$,  the magnetic length can be made to be of the order of $\ell =\sqrt{\hbar/eB}\approx 50nm$. This observation  allows to show the importance of investigating the influence of a disclination in a quantum ring in a  graphene layer, noting that the average  size of the  disclination  in this  carbon material is  of the order of the interatomic distance between two carbon atoms in this nanostructure.   This fact shows that since this defect size is smaller than the magnetic length, it can influence the physical properties for this type of materials (such as quantum Hall effect and other electronic properties), in contrast with common topological defects in which the topological defect is much larger than the magnetic length. In Ref. \cite{akay}  the authors  studied the influence of external magnetic field  for a quantum dot in graphene with presence of  a  topological defect. Recently, the influence  of a disclination  on Landau levels was investigated \cite{Bueno1}. 
By studying a soft confinement in graphene \cite{Sub}, it was observed that the interaction with the substrate of the quantum dots introduces a gap in graphene.  

In this  contribution,   we study  the  two-dimensional quantum ring in  a massive  graphene layer, adopting a model of confining quasiparticle in the ring configuration with the potential proposed   in Ref. \cite{Bakke} based on   using a Dirac oscillator \cite{Moshinsky}  coupling. The Dirac oscillator is a model for a relativistic harmonic oscillator which  in the non-relativistic limit becomed a   harmonic oscillator with a strong spin-orbit coupling. This confining potential  is  a relativistic version of the  Tan-Inkson confining potential in  two-dimensional semiconductor \cite{TanInkson}. In  the original model \cite{Bakke}  two control parameters  were used to obtain a harmonic confinement in a two-dimensional ring. In particular, the quantum point limits are obtained when we make one of the parameters  to be zero, $a_1=0$. In this relativistic model the confining potential is introduced via  the coupling  with the momentum  of  quasiparticle similar to  Dirac oscillator  \cite{Moshinsky}, namely
\begin{equation}
\vec{\mathbf{p}} \rightarrow \vec{\mathbf{p}}  + i \left[ \frac{\sqrt{2Ma_1}}{r} + \sqrt{2Ma_2} r \right] {\gamma}^{0}\widehat{e}_{r} \label{eq1},
\end{equation}
where $a_1$ and $a_2$ are  the characteristic  parameters of the model and   $M$ is a quasiparticle mass. Now if we consider the limit $a_{1}\to 0$, the harmonic confining potential  of the quantum dot is  recovered.
In the case $a_{2} \to 0$, the  relativistic antidot limit is observed.  We use this new coupling in a  Dirac equation to  describe  a quantum ring structure in a massive  graphene in the low-energy limit.   As  it was demonstrated in Ref. \cite{Bakke}, the nonrelativistic limit  of the confining model (\ref{eq1}) reproduces the quantum ring with confining potential proposed by Tan and Inkson \cite{TanInkson}. The  Dirac oscillator coupling  was used in several  applications for  graphene  \cite{strange,jellal,jellal1,jellal2,Bueno,boumeplp}.  Recently, two of us \cite{Amaro} have employed the model proposed in Ref. \cite{Bakke} for  the confining potential and investigated the quantum ring in  a  graphene layer with the presence of  a disclination.

In what follows, we study, in low energy limit, the system described by a massive  Dirac equation, where  a continuous description  near Fermi $K$-points is employed.   We use the Dirac oscillator type coupling to confine harmonically the  quasiparticles  in  a quantum ring pierced by Aharonov-Bohm flux in a disclinated  massive graphene layer submitted  to an uniform magnetic field.  We obtain  the eigenvalues and eigenfunctions which are used to construct the persistent current.  In the case of dynamics in the presence of defects, we  demonstrate the dependence of these physical quantities  on the parameter  characterizing the disclination in a quantum ring.

This paper is organized as follows. In Section II, we study the quantum dynamics of  a quasiparticle in a quantum ring in massive graphene layer in the  presence of an uniform magnetic field.   In Section   III, we obtain the persistent current in the studied system. In Section IV, we discuss spontaneous magnetization and its dependence at zero temperature. And finally, in Section V we present  our conclusions.

\section{ Quantum ring in a  Massive Graphene Layer with Disclinations in the Presence of a  Magnetic Field}\label{sec2}
We can obtain a disclination in  a graphene layer by a procedure known as Volterra process \cite{Moraes}. This transformation can  be represented by a cut  and glue process where we  are cutting and removing/adding a sector in  a  graphene layer, and the resulting disclination is obtained gluing the  new edges of the lips. Due to the symmetry of graphene honeycomb lattice the removed or added angular sector must be a multiple  of $\pi/3$. We can introduce a topological defect in  a graphene layer by  a fictitious gauge field following the approach introduced in Refs. \cite{Lammert,vozcor2,vozcor2,pachos,alex2,Chakraborty,jose}, where a gauge field is introduced in Dirac equation  in order to reproduce the already known effect of the disclination  on the behavior of the spinor \cite{alex2,Lammert}.  In this way we can describe the quantum dynamics of a quasiparticle in a  gapped  graphene layer with   a  disclination by Dirac equation in curved space in the presence of a non-Abelian  gauge field $A_{\mu}^{\pm}$ related with $K$-spin flux. This non-Abelian gauge field is the contribution in the Dirac equation responsible  for mixing of points  $K_{\pm}$ \cite{Lammert,pachos}.  In this way the quantum dynamics of quasiparticles is described by a Dirac equation in  a curved background given by following metric 
\begin{equation}
ds² = dt² - d{r}² - {\alpha}²{r}²d{\varphi}²
 \label{eq30}\end{equation}
where $\alpha$ is the strength  of the disclination  which can be written in terms of the angular sector $\lambda$ which
we removed or inserted in the graphene layer to form the defect, as $ \alpha = 1 \pm \frac{\lambda}{2\pi} $, where in graphene lattice  $\lambda\propto\pi/3$ due to the hexagonal symmetry,  so, we have $\lambda = \pm\frac{N \pi}{3} $, where $N \in [0,6]$ is the number of sectors removed (`-') or added (`+'), that is, positive and negative disclination. 
 A topological defect in  a graphene layer of a disclination type is  described by the line element~(\ref{eq30})   corresponding to removed  sectors or inserted sectors in this honeycomb lattice, the metric (\ref{eq30}) is  a continuous description of   space with removed/inserted angular sector  with a cone geometry. The  presence of  a disclination  in graphene  layer can be  demonstrated in terms of fluxes of fictitious gauge fields through the apex of the graphene cone.  We need  two quantum fluxes to describe the presence of disclination in this medium. The first of these quantum fluxes measures the angular deficit of the cone when a spinor is parallel transported around the apex in a closed path proving thus the variation of the local reference frame along the path.  The flux produced by this geometric contribution acts only on  $A/B$ sublattices and can be determined by using a holonomy transformation \cite{alex2,Lammert,crespi2}, that is
\begin{equation}
\label{paralel2}
\oint \Gamma_{\mu}dx^{\mu}=\pi (\alpha-1)\sigma^{z}\otimes1.
\end{equation}
The second quantum flux is called the K-spin flux, and  it describes the mixing of the Fermi points $K_{+}$ and $K_{-}$ \cite{pachos}. This quantum flux can be written in term of the parameter $\alpha$ as 
\begin{equation}
\label{kspin2}
\oint A^{\pm}_{\mu}dx^{\mu}=-3\pi (\alpha-1)1\otimes\tau^{y}.
\end{equation}
It is important to note that $\tau^{i}$ are the Pauli matrices  acting only on the $K_{\pm}$ space. The (\ref{kspin2}) is  the flux associated  to a non-Abelian gauge field $A^{\pm}_{\mu}$ \cite{Lammert,alex2,pachos}.  The introduction of  the defect  implements a fictitious discontinuity:  when a spinor is  transported  around the apex of disclinated graphene by  a closed path of  $2\pi$ it is forced at some point to jump from a site $B$ to a site $B$ instead of a site $A$ in the sublattices of graphene.  This discontinuity is rectified
by introducing a non-Abelian vector potential $A^{\pm}_{\mu}$ term (\ref{kspin2}) that makes the theory consistent. This contribution is similar  to the Aharonov-Bohm effect for a non-Abelian gauge field.  The defect is induced by the presence of an effective curvature in the structure of graphene, in this way both expressions (\ref{paralel2}) and (\ref{kspin2}) are functions of the parameter $\alpha$ that characterizes the presence of a disclination. Note that, the non-trivial holonomy introduced by phases (\ref{paralel2}) is characterized in the Dirac-Weyl equation by spinor connection and it is introduced  in the Hamiltonian (\ref{hamilton}) due  to curved space description where usual derivative  is changed by the covariant derivative $\partial_{\mu}\longrightarrow \partial_{\mu}-{\Gamma}_{\mu}$.  The effect of the holonomy (\ref{kspin2}) yields a phase given by \cite{Chakraborty,jose}
\begin{equation}
\psi\left(r,\varphi+2\pi\right)=e^{i2\pi\left[\pm\frac{3(\alpha-1)1\otimes\tau^{y}}{2}\right]}\psi\left(r,\varphi\right),
\end{equation}
and  is introduced by a fictitious non-Abelian gauge field,  $A^{\pm}_{\mu}= \pm\frac{3(\alpha-1)1\otimes\tau^{y}}{2r} \hat{e}_{\varphi}$ in the Hamiltonian of the problem. The effective transport of the Dirac spinor around the apex corresponds to a holonomy transformation   for an arbitrary $n$-disclination and represents a non-Abelian Aharonov-Bohm effect due to presence of topological defect. 

In the present paper we have followed the model used in the  papers \cite{vozcor1,alex2,Bueno1,Bueno}, where the influence of disclination  is included in two contributions: the first  one arises from the description of  the Hamiltonian (\ref{hamilton}) in curved space, in this way the  contributions arisen due to holonomy (\ref{paralel2}) are introduced. The second contribution  arising due to the disclination is introduced via a coupling with a non-Abelian field and  is present due to the holonomy (\ref{kspin2}). The confining potential is  implemented via minimal coupling in similar way as in a Dirac oscillator. The external magnetic field  is introduced by a minimal  coupling and has two contributions: one due  to the Aharonov-Bohm flux in the center of the ring, and the other due  to an uniform magnetic field in the ring. 

 Now let us define Dirac matrices representation used in this paper.  We can build the following $4\times4$ representation of Dirac matrices in (2+1)-dimensional space using the $2\times2$ $\tau^{i}$ and $\sigma^{i}$ Pauli matrices  acting  on the Fermi points $K_{\pm}$ and the sublattices labels $A/B$, respectively:
\begin{eqnarray}\label{gammamatrices}
{\gamma}^{0} =\widehat{\beta}&=& {\tau}^{3}\otimes I =\left( \begin{array}{ccc}
I & 0&  \\
0 & -I\end{array}\right), \quad {\gamma}^{i} =\widehat{\beta}\widehat{\alpha}^{i} = i{\tau}^{2}\otimes ({\sigma}^{i}) =    \left( \begin{array}{ccc}
0 & {\sigma}^{i}&  \\
-{\sigma}^{i} & 0\end{array}\right), \nonumber \\ \hat{{\alpha}_i }& =& \left( \begin{array}{ccc}
0 & {\sigma}^{i}&  \\
{\sigma}^{i} & 0\end{array}\right),  \quad \quad \quad \quad \quad \Sigma^{i}=\left(
\begin{array}{cc}
\sigma^{i} & 0 \\
0 & \sigma^{i} \\	
\end{array}\right) ,
\end{eqnarray}
 with ($i = 1,2,3$). $I$ and ${\sigma}^{i}$ stand for $2 \times 2$ the identity  matrix and Pauli matrices respectively and $\vec{\Sigma}$ being the spin vector. The matrices $\widehat{\alpha}^{i}(i = 1,2,3)$ and $\widehat{\beta}$ satisfy the set of properties:
$ \widehat{\alpha}^{i}\widehat{\alpha}^{j} + \widehat{\alpha}^{j}\widehat{\alpha}^{i} = 2{\delta}_{ij}I$, $ \widehat{\alpha}^{i}\widehat{\beta} =- \widehat{\beta} \widehat{\alpha}^{i} $ and ${\widehat{\alpha}^{i2}} =  {\widehat{
\beta} }^{2}=I $. Note that we have used a Dirac-Pauli representation of Dirac matrices \cite{kandemir}. 
 Now we  can write the Dirac equation associated to Hamiltonian (\ref{hamilton}) for gapped graphene with  a  disclination (in a curved space (\ref{eq30})) and considering the presence of non-Abelian gauge field (\ref{kspin2}) due  to the presence of topological defect under the influence of  external magnetic fields and  a confinement potential due to a quantum ring. The result is as follows
\begin{eqnarray}
\left(i{\gamma}^{\mu} \frac{\partial}{\partial x^{\mu}} - i{\gamma}^{\mu}{\Gamma}_{\mu} +{\gamma}^{\mu}  \bold{A}_{\mu}^{C}  - {\gamma}^{\mu} e \bold{A}_{\mu}   - {\gamma}^{\mu}  \bold{A}^{\pm}_{\mu} \right) \Psi = M \Psi.
\label{eq38}
\end{eqnarray}
with $\gamma^{\mu}$ being the Dirac matrices defined in a curved space. The ${\Gamma}_{\mu}$ term is a spinor connection  which is present due to the curved nature of the geometry of the disclinated gapped  graphene lattice  in an elastic  continuous limit (\ref{eq30}). The matrices ${\gamma}^{\mu}$ in a curved space can
be expressed  as functions of triad  fields $e_{a}^{\mu} (x)$. In this curved background frame, the Dirac matrices must
be defined by  $ {\gamma}^{\mu} = e_{a}^{\mu} (x) {\gamma}^{a}$  and satisfy the anticommutation relation $\lbrace {\gamma}^{a},{\gamma}^{b}    \rbrace = 2 {\eta}^{ab}$, where  ${\eta}^{ab}$ the usual Minkowski metric with  a signature ${\eta}^{ab}= diag(+,-,-)$. The $\gamma^{\mu}$ matrices are related to the $\gamma^{a}$ matrices via $\gamma^{\mu}=e^{\mu}_{\,\,\,a}\left(x\right)\gamma^{a}$.
 Moreover, the {\it vierbein fields} satisfy  the relation $ g^{\mu \nu} = e_{a}^{\mu}e_{a}^{\nu} {\eta}^{ab} $.  So that we can write the triad matrix $e_{\mu}^{a}$ and its inverse $e_{a}^{\mu}$ as
 \begin{equation} 
e_{\mu}^{a} = \left( \begin{array}{ccc}
1 & 0 & 0  \\
0 & \cos \varphi &  -\alpha r \sin \varphi \\
0 & \sin \varphi & \alpha r \cos \varphi 
\end{array}\right) , \quad  e_{a}^{\mu} = \left( \begin{array}{ccc}
1 & 0 & 0  \\
0 & \cos \varphi & \sin \varphi \\
0 & -\frac{\sin \varphi}{\alpha r} & \frac{\cos \varphi}{\alpha r}  
\end{array}\right) . \label{eq35}\end{equation}

From matrices (\ref{eq35}), we can obtain the  1-form of connection ${\omega}_{b}^{a}$  through the Maurer-Cartan  structure equation
$de^a +{\omega}_{b}^{a}\wedge e^b = 0 $ (without torsion). Due to the symmetry of the defect,  the connections have only two non- zero components, $
{\omega}_{1}^{2} = - {\omega}_{2}^{1} = -(\alpha - 1)d\varphi
$.  In this way, the spinorial connection  is described in terms of the spin connection, by the equation  $\Gamma_{\mu} =  \frac{i}{4} {\omega}_{\mu \, a\, b} {\Sigma}^{ab}$ such that ${\Sigma}^{ab} = \frac{i}{2} ( {\gamma}^{a}{\gamma}^{b} - {\gamma}^{b}{\gamma}^{a})$.  Hence, the only non-zero spinorial connection component is: 
\begin{equation}
\Gamma_{\varphi} = -\frac{i}{2}(\alpha - 1){\Sigma}³. \label{eq36}
\end{equation}
The  terms contained in Dirac equation have the following origin: the term ${\Gamma}_{\mu} $ is the spinorial connection  arisen due  to the change in the geometry of graphene introduced by disclination. It  is responsible for a non-trivial holonomy due  to the  parallel transport of spinor (\ref{paralel2})  in the geometry  (\ref{eq30}), and  for the variation of the local reference frame in this geometry, and produces a geometric phase that acts in the sublattices $A/B$ of graphene \cite{Lammert,alex2}. The second term $ \bold{A}_{\mu}^{C}$ is the confinement potential given by (\ref{eq1}).  We note  the presence of the term $A^{\pm}_{\mu}= \pm \frac{3}{2r}(\alpha-1)1\otimes\tau^{2}$  due  to the  non-Abelian gauge field related to the K-spin flux   because of the geometric phase (\ref{kspin2}), and contributing  for the mixing of the Fermi points $K_{\pm}$. The coupling with the potential vector  $\bold{A}_{\mu} $ is responsible  for the  inclusion of the Aharonov-Bohm flux\cite{aharonov} piercing through the center of the quantum ring and  an  uniform  magnetic field in the $z$-direction, given by  
\begin{eqnarray}
\vec{\mathbf{A}} =\left[ \frac{\Phi}{2\pi r} + \frac{Br}{2\alpha} \right] \widehat{e}_{\varphi}.\label{eq37},
\end{eqnarray}
 where the first contribution $ \frac{\Phi}{2\pi r}$  arises due  to the  Aharonov-Bohm flux in center of the quantum ring, and the second contribution $ \frac{Br}{2\alpha}$ is due  to an uniform magnetic field in the ring with topological defect in the similar way to Tan-Inkson model \cite{TanInkson,furtado1,furtado2,furtado2} for metallic/semiconductor quantum ring. 
Now, using the triads (\ref{eq35}), we obtain the following equation 

 \begin{align*}
\left[ i{\gamma}^{t} \frac{\partial}{\partial t} + i {\gamma}^{r} \left( \frac{\partial}{\partial r} - \left[ \frac{\sqrt{2Ma_{1}}}{r} - \sqrt{2M a_2} r  \right] {\gamma}^{0}  + \frac{(\alpha - 1) }{2\alpha r}  \right) \right]\Psi  \end{align*} \begin{align} +\left[i{\gamma}^{\varphi} \left( \frac{1}{\alpha r} \frac{\partial}{\partial \varphi} +  i\frac{{\phi}_{AB}}{r} + i\frac{eBr}{2\alpha} + i\frac{a_{\varphi}}{r}      \right) \right]\Psi =  M \Psi, \label{eq39}\end{align} 
 with  $a_{\mu}= \pm \frac{3}{2}(\alpha-1)$, $\phi_{AB} = \frac{\Phi}{ {\Phi}_0}$ and ${\Phi}_0 = \frac{2\pi}{e}$  is the quantum of the magnetic flux. The spin connection is responsible by following term in (\ref{eq39}) $\gamma^{\varphi}\Gamma_{\varphi}=\frac{(\alpha - 1) }{2\alpha r}$ .  We assume the gamma matrices in curved background (\ref{eq30})  to be given by:

 \begin{eqnarray}
{{\gamma}^{\mu}} = e_{a}^{\mu} {\gamma}^{a} = \left\{ \begin{array}{ll}
{\bar{\gamma}}^{0} = {\gamma}^{0} = {\gamma}^{t};   & \\
{\bar{\gamma}}^{1} = {\gamma}^{1} cos \varphi
+  {\gamma}^{2}sen \varphi = {\gamma}^{r} ; \\
 {\bar{\gamma}}^{2} = - \frac{{\gamma}^{1}}{\alpha r} sen \varphi
+  \frac{{\gamma}^{2}}{\alpha r}cos \varphi = \frac{{\gamma}^{\varphi} }{\alpha r} .
\end{array} \right.
\end{eqnarray}

From (\ref{eq39}), using the gamma matrices (\ref{gammamatrices}) we can  rewrite  the effective Hamiltonian for this system in the following form

\begin{eqnarray}
H = - i{\hat{\alpha}}^{r}  \left( \frac{\partial}{\partial r} - \left[ \frac{\sqrt{2Ma_{1}}}{r} - \sqrt{2M a_2} r  \right] \hat{\beta}  + \frac{(\alpha - 1) }{2\alpha r}  \right) \nonumber \\  -i\hat{\alpha}^{\varphi} \left( \frac{1}{\alpha r} \frac{\partial}{\partial \varphi} +  i\frac{{\phi}_{AB}}{r} + i\frac{eBr}{2\alpha} + i\frac{a_{\varphi}}{r}\right) + \hat{\beta} M ,\end{eqnarray}
where we define $ a_{\varphi}=\pm \frac{3}{2}(\alpha-1)$.   Now the Dirac equation is given by $H\Psi=E\Psi$.   Splitting this equation into two decoupled  equations is well known  in the massless case \cite{vozmediano,castro,Lammert,crespi2} but this  is  possible also in some  situations in  a massive case \cite{ruegg,jose,Chakraborty2}. In the present  case a transformation  cannot be defined  from results for an effective Abelian gauge field. In this way we use the non-Abelian phase (9) in the construction of the effective Hamiltonian.  
We use a similarity transformation  to eliminate the contribution arisen due to the spinorial connection in the Dirac equation (\ref{eq39}) \cite{Villa}. We  choose the following similarity transformation   $S(\varphi) = e^{-i\frac{\varphi}{2} {\Sigma}^{3}}$  to change the representation of local Dirac matrices, $({\gamma}^{r} , {\gamma}^{\varphi} )= {\bar{\gamma}}^{j}$. Thus, we  write  the  Ansatz for $\Psi'$ given by
\begin{eqnarray}
\Psi' = e^{-iE t -i\frac{\varphi}{2}{\Sigma}³}\left(\begin{array}{c} \psi(r,\varphi) \\  \chi (r,\varphi) \end{array}\right) 
\label{eq40}
\end{eqnarray} 
where, as  it was mentioned above, $ \psi =\psi(r,\varphi)$ and $  \chi = \chi (r,\varphi)$ represent the sublattices $ A$ and $B$ respectively. From the similarity transformations  and Ansatz (\ref{eq40}), we obtain the  set of   equations:
\begin{eqnarray} 
(E - M)\psi = -i{\sigma}¹\left[ \frac{\partial }{\partial r} + \frac{1}{2r}   +  \frac{\sqrt{2Ma_{1}}}{r} + \sqrt{2M a_2} r  \right]\chi  \nonumber\\ -i{\sigma}²\left[  \frac{1}{\alpha  r} \frac{\partial}{\partial \varphi}  +\frac{i {\phi}_{AB}}{r}  +\frac{ieBr}{2\alpha}+ \frac{{i a}_{\varphi}}{r} \right]\chi 
\label{eq41}
\end{eqnarray}
and
\begin{eqnarray} 
(E+ M)\chi = -i{\sigma}¹\left[ \frac{\partial }{\partial r} + \frac{1}{2r}   - \frac{\sqrt{2Ma_{1}}}{r} - \sqrt{2M a_2} r  \right]\psi \nonumber\\ -i{\sigma}²\left[  \frac{1}{\alpha  r} \frac{\partial}{\partial \varphi}  +\frac{i {\phi}_{AB}}{r}  +\frac{ieBr}{2\alpha}+ \frac{{i a}_{\varphi}}{r} \right]\psi .
\label{eq42}
\end{eqnarray}
Now, we  eliminate the  $ \chi$ in (\ref{eq41}) substituting  (\ref{eq42})  into it, and obtain the following  second-order differential equation:
 \begin{eqnarray}  {(E² - M²) \psi} &= & -  {\frac{{\partial}^2 {\psi}}{\partial {r}²} -\frac{1}{r}\frac{\partial \psi}{\partial r} +\frac{\psi}{4{r}²} - \frac{\sqrt{2Ma_1}}{{r}²}\psi 
+ \sqrt{2Ma_2}\psi  + \frac{2Ma_1}{{r}²}\psi} + 2Ma_2 {r}² \psi +\nonumber\\&+& 4M\sqrt{a_1 a_2}\psi  -  \frac{1}{{ \alpha}^2 {r}²}\frac{{\partial}²\psi}{\partial{\varphi}²}+
i\frac{{\sigma}³}{\alpha{r}²}\frac{\partial \psi}{\partial \varphi}- 2i{\sigma}³\frac{\sqrt{2Ma_1}}{\alpha r²}\frac{\partial \psi}{\partial \varphi}-  \frac{2i{\sigma}³\sqrt{2Ma_2}}{\alpha}\frac{\partial \psi}{\partial \varphi} -\nonumber\\&-& 2i\frac{{\phi}_{AB}}{\alpha r²}\frac{\partial \psi}{\partial \varphi} - \frac{iBe}{{\alpha}^2}\frac{\partial \psi}{\partial \varphi}  - \frac{2i a_{\varphi}}{\alpha r²} \frac{\partial \psi}{\partial \varphi}   \nonumber \\ &+& 2 eB{\sigma}³ \frac{\sqrt{2 {Ma_1 }}}{2\alpha} \psi + 2{\sigma}^3{a}_{\varphi} \frac{\sqrt{2M{a}_1}}{r²} \psi +2{\sigma}³ {{\phi}_{AB}} \frac{\sqrt{2Ma_1}}{{r}²}\psi  + 2{\sigma}³{{\phi}_{AB}} \sqrt{2Ma_2} \psi  \nonumber \\ 
&+& 2eB{\sigma}³ \frac{\sqrt{2Ma_2 }  r²}{2\alpha} \psi + 2{\sigma}^3\sqrt{2Ma_2}{a}_{\varphi}\psi      +  {\sigma}³\frac{eB }{2 \alpha}\psi      -{\sigma}^3\frac{{a}_{\varphi}}{r²} \psi -{\sigma}³\frac{{\phi}_{AB}}{r²}\psi \nonumber\\ &+ & \frac{{{\phi}_{AB}}²}{r²} \psi + 2eB\frac{ \ {\phi}_{AB}}{2\alpha}\psi +\frac{{{a}_{\varphi}}^2}{r²}\psi  + \frac{2{\phi}_{AB} \ a_{\varphi}}{r²}\psi + 2eB\frac{ a_{\varphi}}{2\alpha}\psi + \frac{e² B² r²}{4{\alpha}²}\psi .
\label{eq43}\end{eqnarray} 
 Then, to obtain the solution of  the Eq.(\ref{eq43}), we make  the following  choice for the  function $\psi$:
\begin{equation}
\psi (r, \varphi) =  e^{im\varphi} \left(\begin{array}{c}R_+ (r) \\ R_{-} (r)  \end{array}\right) = e^{im\varphi}R_s (r)  ,
\end{equation} 
where $m= l + \frac{1}{2}$, is semi-integer  with $l = 0,\pm1,\pm2,...$ and $R_{s} = \left(R_{+}(r) , R_{-}(r) \right)$.  In this way we obtain the following set of equations
 \begin{equation} 
\left[ \frac{d²}{d{r}²} + \frac{1}{r}\frac{d}{dr} - \frac{{{\delta}_s}²}{{\alpha}²{r}²} - \frac{{{\omega}_{\alpha}}² M²}{4}{{r}²}+{\epsilon}_s \right]R_s (r) = 0\label{eq45},
\end{equation} 
with parameters ${\delta}_s$, ${\vartheta}_{s}$, ${\epsilon}_s$ and   ${\omega}_{\alpha}$ defined as:
\begin{eqnarray}  
{\delta}_s &=& {\vartheta}_s + \alpha s\sqrt{2Ma_1}  ,\nonumber \\   {\vartheta}_s &=&  (l + \alpha{\phi}_{AB} )  + \frac{(1- \alpha s)}{2} + \alpha {a}_{\varphi}, \nonumber\\
{\epsilon}_s &=&  {E}² - M² - M\frac{({\delta}_s + \alpha s)}{\alpha}\left({\omega}_0 s  +  \frac{{\omega}_c}{\alpha} \right), \nonumber\\ {\omega}_{\alpha} &=& \sqrt{ {{\omega}_0}²   +  \frac{{2{\omega}_0}{\omega}_c s}{\alpha} + \frac{{{\omega}_c}²}{{\alpha}²} } \label{eq46}
\end{eqnarray}
with $\omega_{0} =\sqrt{\frac{8a_{2}}{M}}$  being the characteristic frequency and $\omega_{c} =\frac{eB}{M} $ is the cyclotron frequency. To solve the equation (\ref{eq45}) we make the change of the variable $ \rho = \sqrt{\frac{{{\omega}_{\alpha}}²M²}{4}}{r}²  $  so that  afterwards our equation takes the form:
\begin{equation} 
\left[ \rho \frac{d²}{d{\rho}²} + \frac{d}{d\rho} - \frac{{{\delta}_s}²}{4{\alpha}^2 \rho} - \frac{{\rho}}{4} +\frac{{\epsilon}_s}{2M{\omega}_{\alpha}} \right]R_s (\rho) = 0. \label{eq47}\end{equation}
In this way, we do the asymptotic analysis of Eq. (\ref{eq45}), for the limits $R_s \rightarrow 0$ and $r \rightarrow \infty$, it is possible to present radial equation in the following form:
\begin{equation}
R_s(\rho)= e^{\frac{-\rho}{2}}{\rho}^{\frac{|{\delta}_s|}{2\alpha}}F_s (\rho)
\label{eq48}
\end{equation}
Substituting this into the previous equation we obtain:
\begin{equation} 
 \rho \frac{d²F_s(\rho)}{d{\rho}²} + \left[ \frac{|{\delta} _{s}|}{\alpha} + 1 - \rho \right] \frac{dF_s(\rho)}{d \rho}  + \left[\frac{{\epsilon}_s}{2M{\omega}_{\alpha}} -\frac{|{\delta}_s|}{2\alpha} - \frac{1}{2} \right]F_s (\rho) = 0 \label{eq49},
 \end{equation}
 where Eq. (\ref{eq49}) is the hypergeometric equation whose solution is the hypergeometric function:
 \begin{equation}  
   F_s (\rho) =  {\phantom{1}_1}F_1(a,b;z)=
 {\phantom{1}_1}F_1\left(\frac{|{\delta}_s|}{2\alpha} + \frac{1}{2} -\frac{{\epsilon}_s}{2M{\omega}_{\alpha}},\frac{|{\delta}_s|}{\alpha}+1,\rho \right)\label{eq50}. 
 \end{equation}
Now to get a finite solution anywhere, we impose the condition  that the solution of hypergeometric series becomes a polynomial of degree $n$, that is:
 $\frac{|{\delta}_s|}{2\alpha} + \frac{1}{2} -\frac{{\epsilon}_s}{2M{\omega}_{\alpha}} = -n
 \label{eq57}$
 from this equation, substituting the parameters (\ref{eq46}), we obtain the energy spectrum for the particle confined in two-dimensional quantum ring  pierced by Aharonov-Bohm quantum flux in gapped  graphene  in the presence  of the uniform  magnetic field:
\begin{equation} 
{{E}²}_{n,l}  = \left( 2n + \frac{\vert{\delta}_{s} \vert}{{\alpha}} + 1  \right) M{\omega}_{\alpha} +  M\frac{({\delta}_s + \alpha s)}{\alpha}\left({\omega}_0 s  +  \frac{{\omega}_c}{\alpha} \right) +  M²  ,\label{eq51}    
\end{equation}
where $s = +1$ corresponds to sublattice $A$ and $s = -1$ corresponds to sublattice $B$.
Note that the spectrum depends on the quantum numbers $n$ and $l$, the  cyclotron frequency ${\omega}_c $, and, in addition, on the new frequency related to the topological  defect ${\omega}_{\alpha}$  through the dependence on the parameter $\alpha$.  Note that ${\delta}_s = (l + \alpha{\phi}_{AB} )  + \frac{(1- \alpha s)}{2} + \alpha {a}_{\varphi} + \alpha s\sqrt{2Ma_1}  $ where $\phi_{AB}=\frac{\Phi}{\Phi_{0}}$.  This way the eigenvalues of energy  (\ref{eq51}) are periodic functions of $\Phi$ with the period $\Phi_{0}$. A similar behaviour was observed for semiconductor quantum ring in Refs.\cite{TanInkson,furtado3} . Moreover, the Eq. (\ref{eq51}) has a contribution of the non-Abelian gauge field ${A}_{\varphi}^{\pm}$  responsible for  the Aharonov-Bohm effect of the spinors, due to the presence of  the  disclination.

\begin{figure}[!htb]\label{figura1}
\includegraphics[scale=1.0]{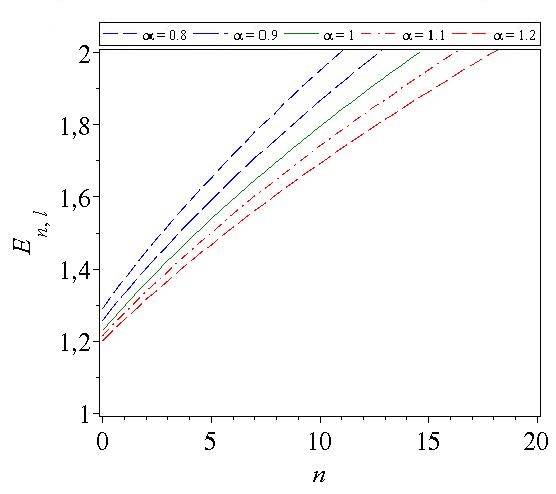} 
\caption{The obtained energy spectrum as a function of  quantum number $n$ for several values of $\alpha$. We have adopted the  following values for model parameters: $B = 1 T$, $M = 1$, $l = 1$, ${\phi}_{AB} = 1$, and $s=1$. $a_1 = 9.0122\times 10^{-6} eV.m^{2}$ and $ a_2 =  2.222\times 10^{-5} eV.m^{-2}$ are the Tan -Inkson parameters \cite{TanInkson}.  }
\end{figure}

Notice that according to  Fig. 1, for the positive disclinations the energy eigenvalues have magnitude  greater than  in the flat case, while  in the case of negative disclinations the energy assumes smaller values for the same quantum number $n$.

The solutions with positive energy describe the dynamics of electrons  in the conduction band while   negative energy corresponds to the dynamics of holes in the valence  band. In this paper we address only the first case. Also, as it was argued earlier, components of the spinor in graphene describe the contributions of sub-lattices, with the real  spins of the particles are not considered within this approach. For equation (\ref{eq51}), the spinors corresponding to positive energy  for $s= +1$ and ${\eta}_{-}(r) = 0 $ are written as:
\begin{eqnarray}
&  \Psi_+ =  f_+ {\phantom{1}_1}F_1 \left(-n,\frac{|{\delta}_+|}{\alpha}+1, \frac{{\omega}_{\alpha} M}{2}{r}²\right) \times \nonumber \\ &
\left( \begin{array}{c} 1 \\ 0 \\ 0 \\  \frac{i}{E+M}\left[  \frac{M}{2} [ ({\omega}_0  + {\omega}_c /\alpha  ) + {\omega}_0  ]r + \frac{\left( {\vartheta}_+ - |{\delta}_+| + \alpha \sqrt{2Ma_1} \right) }{\alpha r} - \frac{M{\omega}_c r}{2\alpha}  \right]
\end{array} \right)\nonumber \\&+  \frac{i f_+}{E+M}\left(\frac{n M({\omega}_0  + {\omega}_c /\alpha )r}{           \frac{|{\delta}_+|}{\alpha}+1} \right){\phantom{1}_1}F_1 \left(-n + 1 \ ,\frac{|{\delta}_{+}|}{\alpha}+2 \ , \frac{{\omega}_{\alpha}M}{2}{r}²\right)\left( \begin{array}{c} 0 \\ 0 \\ 0 \\ 1\end{array} \right), \label{eq52}  \end{eqnarray}

and  for $s= -1$ and ${\eta}_{+}(r) = 0 $ as

\begin{eqnarray}
&  \Psi_{-} =  f_{-} {\phantom{1}_1}F_1 \left(-n,\frac{|{\delta}_{-}|}{\alpha}+1, \frac{{\omega}_{\alpha}M}{2}{r}²\right) \times \nonumber \\ &
\left( \begin{array}{c} 0 \\ 1 \\ \frac{i}{E+M}\left[  \frac{M}{2} [ ({\omega}_0  + {\omega}_c /\alpha  ) + {\omega}_0  ]r - \frac{\left( {\vartheta}_{-} + |{\delta}_{-}| - \alpha \sqrt{2Ma_1} \right) }{\alpha r} + \frac{M{\omega}_c r}{2\alpha}  \right] \\ 0 
\end{array} \right)\nonumber \\&+   \frac{i f_{-}}{E+M}\left(\frac{nM({\omega}_0  + {\omega}_c /\alpha )r}{           \frac{|{\delta}_{-}|}{\alpha}+1} \right){\phantom{1}_1}F_1 \left(-n + 1 \ ,\frac{|{\delta}_{-}|}{\alpha}+2 \ , \frac{{\omega}_{\alpha}M}{2}{r}²\right)\left( \begin{array}{c} 0 \\ 0 \\ 1 \\ 0\end{array} \right). \label{eq53} \end{eqnarray}

 Both in (\ref{eq52}) and (\ref{eq53}), the factor $f_s = N_s e^{-iEt}$  reads

\begin{eqnarray} 
 {f}_s =  \left( \frac{M {\omega}_{\alpha}  \Gamma \left(\frac{|{\delta}_s|}{\alpha}+n+1 \right)}{\Gamma(n+1)\left[\Gamma \left(\frac{|{\delta}_s|}{\alpha}+1\right)\right]²} \right)^{1/2}  \times e^{-iEt}e^{i(l+1/2)\varphi} e^{- \frac{{\omega}_{\alpha} M}{4}{r}^{2}} \left(\frac{{{\omega}_{\alpha}}² M²}{4} \right)^{\frac{|{\delta}_s|}{4\alpha}} {r}^{\frac{|{\delta}_s|}{\alpha}}.
 \label{eq54}\end{eqnarray} 
 
Note that at  $\alpha = 1$, the non-Abelian gauge field  $\alpha_{\phi}$ goes to zero and the parameters  ${\delta}_s$, ${\vartheta}_{s}$, ${\epsilon}_s$ and   ${\omega}_{\alpha}$   tend to the values of the parameters ${\lambda}_s $, ${\xi}_s $, ${\varepsilon}_s $ and $\omega $  respectively. In other words, we recover the spectrum, persistent current, and spinors of positive energy for the case of flat sheet without defect, i.e., $\alpha = 1$ represents  the absence of topological defect of the gapped  graphene layer.  Moreover, if we  switch off  the magnetic field, i.e.,  make $B=0$ and by consequence ${\omega}_c = 0$, we recover the case of  a two-dimensional ring in   a gapped graphene layer, and in the case  $M=0$ we obtain the results  which  were considered in Ref. \cite{Amaro}.

\section{The persistent current in the presence of a disclination }\label{sec3}
 The study   of small  conducting rings  pierced by a magnetic flux in low temperature limit  predicted the emergence of  persistent  currents \cite{buttiker} in these systems.  This current is a thermodynamic effect  deeply connected with the presence of quantum coherence. The  theoretically predicted persistent current  has been   experimentally observed in a ring  in several  experiments
in metals \cite{51,52,53} and semiconductors \cite{54}.
Now let us calculate the persistent current for our model of quantum ring  in a graphene layer with a disclination. The persistent current carried by a given electron state is calculated using the Byers-Yang relation~\cite{Byers}. This relation expresses the persistent current as a derivative of the energy with respect to the magnetic flux 
\begin{eqnarray}\label{current2}
I =-\frac{\partial U}{\partial \Phi}= -\sum_{n,l} \frac{\partial E_{n,l}}{\partial \Phi},
\end{eqnarray}
 where $U=\sum_{n,l}E_{n,l}$ is total energy of system and  we consider the zero temperature $T=0$ and a fixed number of quasiparticles $n$. The summation is realized over the $n$  occupied states.

In this way,  the persistent current for our system reads:
\begin{eqnarray} 
I &=& \left( \frac{\mp e}{ 4\pi} \right)\sum_{n,l} M  \left(\frac{{\delta}_s}{\vert {\delta}_s   \vert} {\omega}_{\alpha} + \frac{{\omega}_{c}}{\alpha}  + s{\omega}_0 \right) \times \nonumber\\&\times&
 \Bigg\{ M² +  M\frac{({\delta}_s + \alpha s)}{\alpha}\left({\omega}_0 s  +  \frac{{\omega}_c}{\alpha} \right) + \left( 2n + \frac{\vert{\delta}_{s} \vert}{{\alpha}} + 1  \right) M{\omega}_{\alpha} \Bigg\}^{-1/2} 
\label{eq56}\end{eqnarray} 

Note that the  persistent current is a function of the parameter characterizing  the curved geometry introduced by  a topological defect. The expression for  the persistent current given by  (\ref{eq56}) is a function of the non-Abelian magnetic flux and the Aharonov-Bohm flux  on which the parameter ${{\delta}_s}$ depends. The current has jumps with a certain periodicity and is function of the  Aharonov-Bohm flux $\Phi$. Fig.2 shows the predicted behavior when the  oscillations vanish for the large values of the Aharonov-Bohm flux.

\begin{figure}[!htb]\label{fig2}
\includegraphics[scale=1.0]{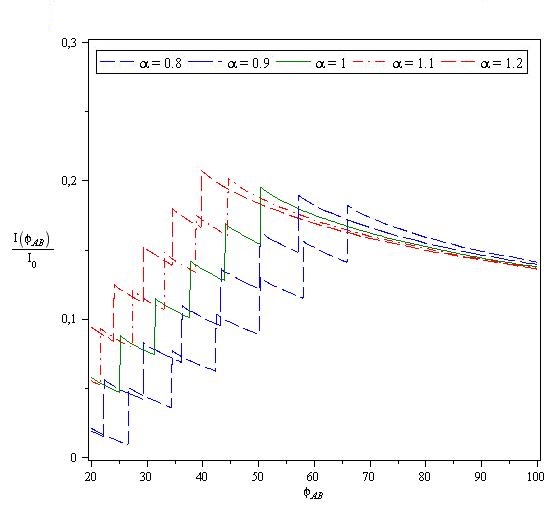} 
\caption{The behavior of persistent current as a function of magnetic flux ${\phi}_{AB}=\frac{\Phi}{\Phi_0}$ for several values of $\alpha$. The  topological current term is defined as $I_0 = \frac{1}{2{\phi}_0} \left(  \sqrt{\frac{8a_2}{M}}s + \frac{eB}{\alpha M} \right)$. The values for the other parameters are  $B = 1 T$, $M = 1$, $s=1$, $ -10 \leq l \leq 10 $ and $ 0\leq n \leq 5$. $a_1 = 9.0122\times 10^{-6} eV.m^{2}$ and $ a_2 =  2.222\times 10^{-5} eV.m^{-2}$ }. 
\end{figure}
We can compare  our results with those obtained for quantum dots in graphene. The predicted here oscillatory behavior  turns out to be similar  to the one observed in Ref. \cite{Bueno1}.

\section{The  magnetization for quantum ring in presence of Disclination }\label{sec4}
Now, we investigate the magnetization for a quasiparticle in quantum ring in  a graphene sheet at zero temperature. In general,  an analytic expression for a magnetization $\mu$  is defined via applied magnetic field $B$ as follows  
\begin{eqnarray}
 \mu(B)= - \frac{\partial U}{\partial B} =- \sum_{n,l}  \frac{\partial {E}_{n,l}}{\partial B} ,
\end{eqnarray} 
where we consider a system with a temperature $T = 0$ and a fixed number $N$ of spinless particles and the summation is realized over the $N$ occupied states. Using (\ref{eq51}), we obtain:
\begin{equation}
\mu(B,\alpha) = - \sum_{n,l} \left(\frac{s e}{2}\right) {  \left[ \frac{\left( 2n + \frac{|{\delta}_s |}{\alpha}+ 1 \right)}{\alpha}  + \frac{{\delta}_s + \alpha s}{{\alpha}²}\right]   } \times \end{equation} 
\begin{equation}
\Bigg\{ M² +  M\frac{({\delta}_s + \alpha s)}{\alpha}\left({\omega}_0 s  +  \frac{{\omega}_c}{\alpha} \right) + \left( 2n + \frac{\vert{\delta}_{s} \vert}{{\alpha}} + 1  \right) M{\omega}_{\alpha} \Bigg\}^{-1/2} 
\label{eq58},\end{equation} 

\begin{figure}[!htb]\label{figura3}
\includegraphics[scale=0.5]{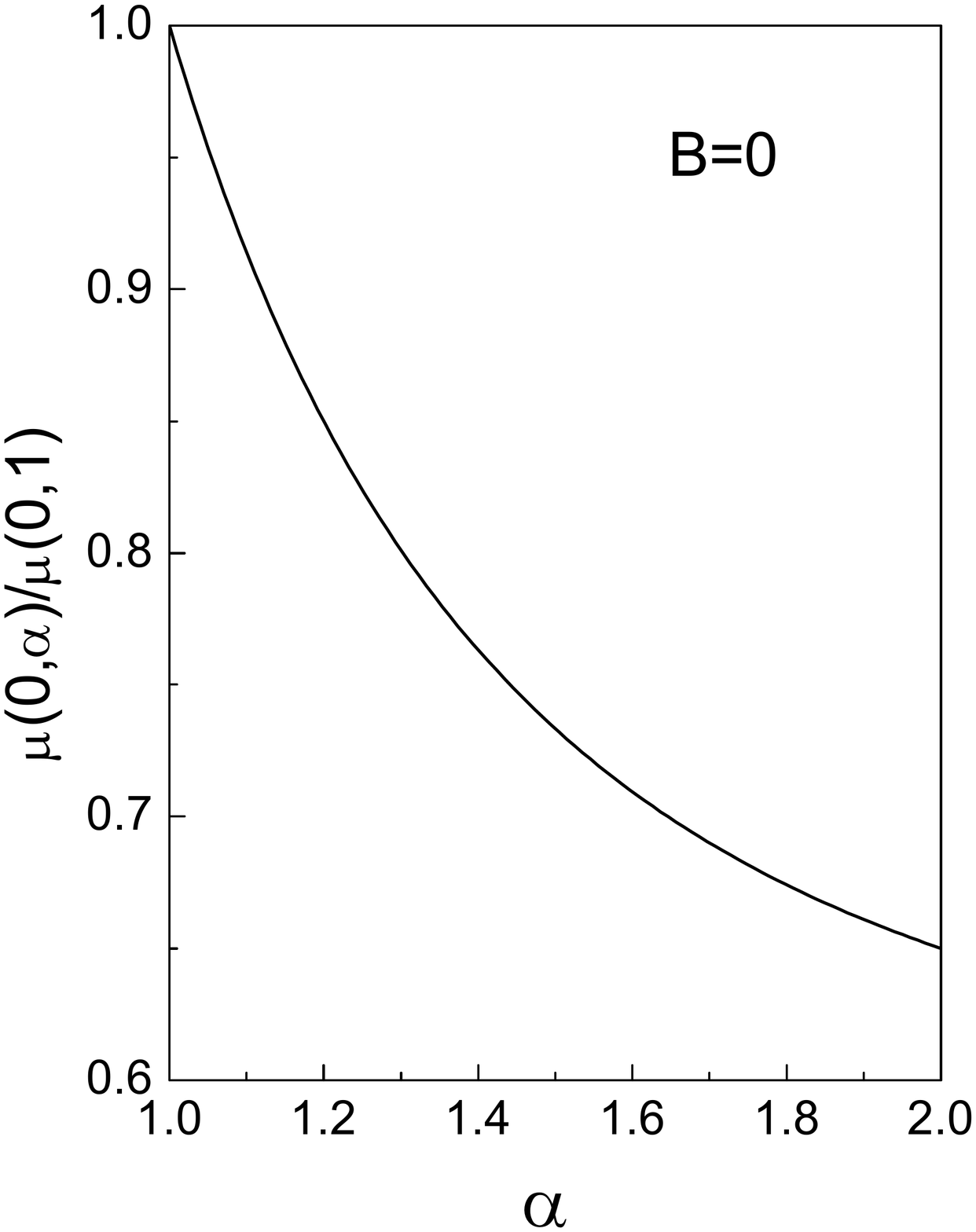} 
\caption{The behavior of normalized zero-field (spontaneous) magnetization as a function of disclination parameter $\alpha$. In this plot, we used $B=0$, $M = 1$, ${\phi}_{AB} = 1$, $s= -1$, $a_1 = 9.0122\times 10^{-6} eV.m^{2}$ and $ a_2 =  2.222\times 10^{-5} eV.m^{-2}$. The sum runs over  $ -10 \leq l \leq 10 $ and $ 0\leq n \leq 5$. }
\end{figure}

\begin{figure}[!htb]\label{figura4}
\includegraphics[scale=0.5]{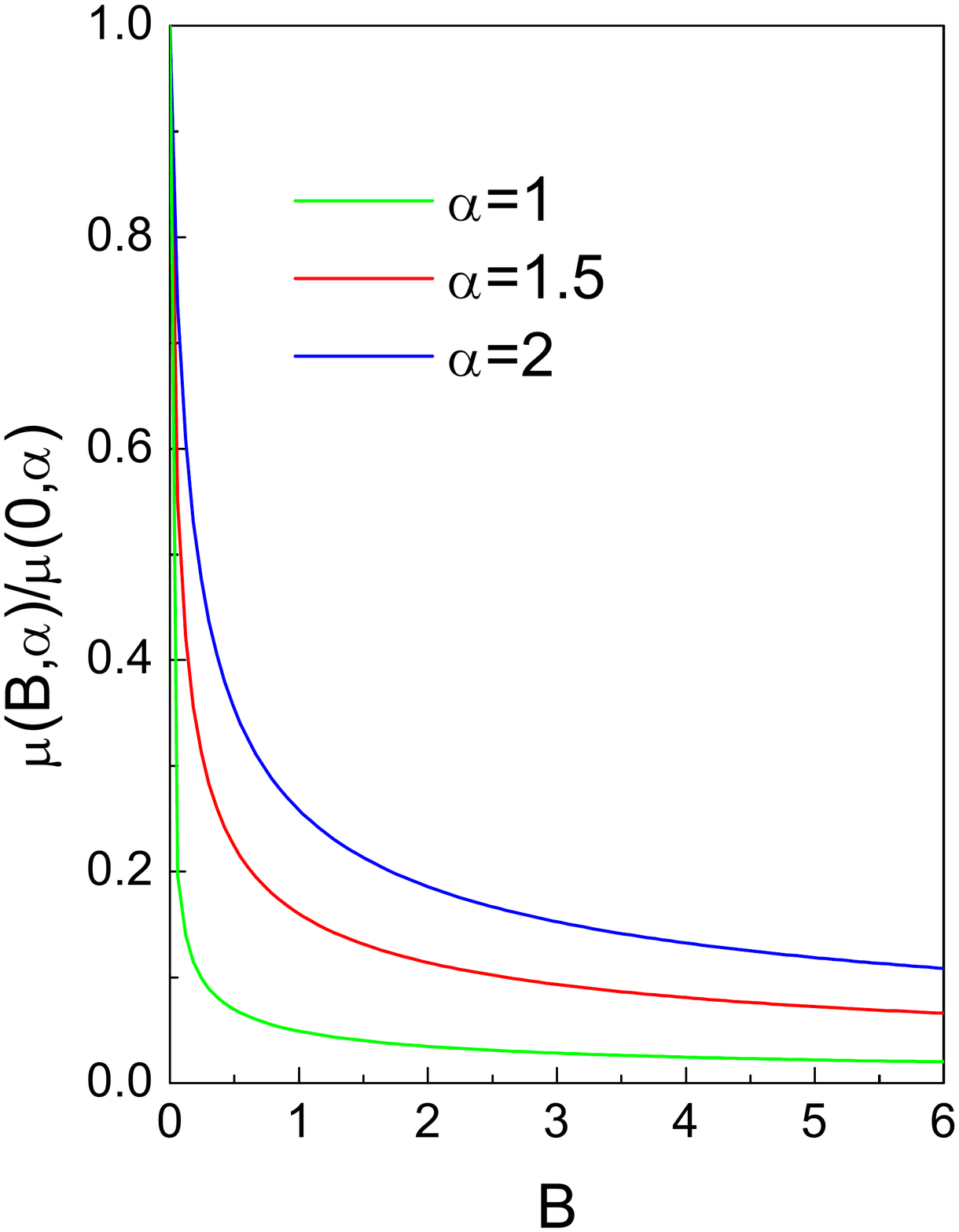} 
\caption{Magnetic field dependence of normalized magnetization for different values of disclination parameter $\alpha$ changing from defect-free value of $\alpha=1$ (corresponding to $N=0$) to its maximum number $\alpha=2$ (corresponding to $N=6$). In this plot, we used $M = 1$, ${\phi}_{AB} = 1$, $s= -1$, $a_1 = 9.0122\times 10^{-6} eV.m^{2}$ and $ a_2 =  2.222\times 10^{-5} eV.m^{-2}$. The sum runs over  $ -10 \leq l \leq 10 $ and $ 0\leq n \leq 5$. }
\end{figure}

We can conclude from Eq (\ref{eq58}) that the spontaneous zero-field magnetization strongly depends on the parameter $\alpha$  characterizing the presence of disclination in a graphene layer. This dependence is shown in Fig. 3  where $\mu(0,1)$ corresponds to spontaneous magnetization of the defect-free system with $\alpha=1$. And finally, Fig.4 depicts field dependence of the induced magnetization $\mu (B,\alpha)$ for different values of the defect controlling paramter $\alpha$. Observe that high field magnetization quite markedly increases with the number of defects in the studied system. 
\section{Conclusions}\label{sec5}
In this contribution we have studied a quantum ring in a gapped graphene layer with  a topological defect submitted  to an uniform magnetic field  along $z$-direction. We use a confinement potential proposed in Ref. \cite{Bakke} to confine a quasiparticle in  a two-dimensional quantum ring. This potential was introduced in Dirac equation  using minimal coupling similar to Dirac oscillator. We have considered an Aharonov-Bohm magnetic flux in the center of quantum ring. The quantum dynamics  of the massive quasiparticle in the presence of  a disclination  is determined by a quantum ring potential. The eigenvalues of energy are obtained, and we found that  the   energy is  a periodic function of  the Aharonov-Bohm flux $\Phi$ with the period $\Phi_{0}$. We have  demonstrated the  influence of  the disclination on energy levels. Namely, for  a positive disclination the energy eigenvalues increase  with growing of parameter of disclination $\alpha$, while for the case  of a negative disclination $\alpha>1$ eigenvalues of energy decrease  in comparison with  the case of a gapped graphene layer without  a  defect. The  persistent current, for  $T=0$, was obtained, and its expression  turns out to depend on  the parameter $\alpha$  characterizing the presence of the disclination.   It  has jumps with a certain periodicity  on the non-Abelian field  $a_{\varphi}=\pm\frac{3}{2}(\alpha-1)$ and the Aharonov-Bohm $\phi_{AB}=\frac{\Phi}{\Phi_{0}}$ fluxes. We have also obtained an analytic expression for zero-temperature magnetization and  found that it is a periodic function of fluxes. We conclude that the role of geometry introduced by topological defects in the electronic properties of graphene is central for the realization of the future investigation  regarding  quantum computation  in these systems \cite{Knut05,knut5,knut8}.

{\bf Acknowledgments.}  We are grateful to Knut Bakke for interesting discussion. We thank  CNPq, CAPES, CNPq/Universal and FAPESQ  for financial support.

\end{document}